\documentclass[superscriptaddress,onecolumn,secnumarabic,
amssymb,amsmath,nobibnotes,aps,prd,showkeys,showpacs,nofootinbib]{revtex4}

\usepackage[latin1]{inputenc}
\usepackage{graphicx}
\usepackage[english]{babel}
\usepackage{slashed}
\usepackage{amsmath}
\usepackage{amssymb}
\usepackage{amsfonts}
\usepackage{colordvi}
\usepackage{psfrag}
\usepackage{color}

\begin{document}
\def\cL{{\cal L}}
\def\be{\begin{equation}}
\def\ee{\end{equation}}
\def\bea{\begin{eqnarray}}
\def\eea{\end{eqnarray}}
\def\beq{\begin{eqnarray}}
\def\eeq{\end{eqnarray}}
\def\tr{{\rm tr}\, }
\def\nn{\nonumber \\}
\def\e{{\rm e}}


\def\bef{\begin{figure}}
\def\eef{\end{figure}}
\newcommand{\ans}{ansatz }
\newcommand{\eeqn}{\end{eqnarray}}
\newcommand{\bd}{\begin{displaymath}}
\newcommand{\ed}{\end{displaymath}}
\newcommand{\mat}[4]{\left(\begin{array}{cc}{#1}&{#2}\\{#3}&{#4}
\end{array}\right)}
\newcommand{\matr}[9]{\left(\begin{array}{ccc}{#1}&{#2}&{#3}\\
{#4}&{#5}&{#6}\\{#7}&{#8}&{#9}\end{array}\right)}
\newcommand{\matrr}[6]{\left(\begin{array}{cc}{#1}&{#2}\\
{#3}&{#4}\\{#5}&{#6}\end{array}\right)}
\newcommand{\cvb}[3]{#1^{#2}_{#3}}
\def\lsim{\raise0.3ex\hbox{$\;<$\kern-0.75em\raise-1.1ex
e\hbox{$\sim\;$}}}
\def\gsim{\raise0.3ex\hbox{$\;>$\kern-0.75em\raise-1.1ex
\hbox{$\sim\;$}}}
\def\abs#1{\left| #1\right|}
\def\simlt{\mathrel{\lower2.5pt\vbox{\lineskip=0pt\baselineskip=0pt
           \hbox{$<$}\hbox{$\sim$}}}}
\def\simgt{\mathrel{\lower2.5pt\vbox{\lineskip=0pt\baselineskip=0pt
           \hbox{$>$}\hbox{$\sim$}}}}
\def\unity{{\hbox{1\kern-.8mm l}}}
\newcommand{\eps}{\varepsilon}
\def\ep{\epsilon}
\def\ga{\gamma}
\def\Ga{\Gamma}
\def\om{\omega}
\def\omp{{\omega^\prime}}
\def\Om{\Omega}
\def\la{\lambda}
\def\La{\Lambda}
\def\al{\alpha}
\newcommand{\ov}{\overline}
\renewcommand{\to}{\rightarrow}
\renewcommand{\vec}[1]{\mathbf{#1}}
\newcommand{\vect}[1]{\mbox{\boldmath$#1$}}
\def\tm{{\widetilde{m}}}
\def\mcirc{{\stackrel{o}{m}}}
\newcommand{\Dm}{\Delta m}
\newcommand{\dm}{\varepsilon}
\newcommand{\tanb}{\tan\beta}
\newcommand{\nbar}{\tilde{n}}
\newcommand\PM[1]{\begin{pmatrix}#1\end{pmatrix}}
\newcommand{\up}{\uparrow}
\newcommand{\down}{\downarrow}
\def\omE{\omega_{\rm Ter}}
%

\newcommand{\Dsusy}{{susy \hspace{-9.4pt} \slash}\;}
\newcommand{\DCP}{{CP \hspace{-7.4pt} \slash}\;}
\newcommand{\mc}{\mathcal}
\newcommand{\gr}{\mathbf}
\renewcommand{\to}{\rightarrow}
\newcommand{\gtc}{\mathfrak}
\newcommand{\wh}{\widehat}
\newcommand{\br}{\langle}
\newcommand{\kt}{\rangle}


\def\lsim{\mathrel{\mathop  {\hbox{\lower0.5ex\hbox{$\sim$}
\kern-0.8em\lower-0.7ex\hbox{$<$}}}}}
\def\gsim{\mathrel{\mathop  {\hbox{\lower0.5ex\hbox{$\sim$}
\kern-0.8em\lower-0.7ex\hbox{$>$}}}}}

\def\nn{\\  \nonumber}
\def\de{\partial}
\def\brf{{\mathbf f}}
\def\bbf{\bar{\bf f}}
\def\bF{{\bf F}}
\def\bbF{\bar{\bf F}}
\def\bA{{\mathbf A}}
\def\bB{{\mathbf B}}
\def\bG{{\mathbf G}}
\def\bI{{\mathbf I}}
\def\bM{{\mathbf M}}
\def\bY{{\mathbf Y}}
\def\bX{{\mathbf X}}
\def\bS{{\mathbf S}}
\def\bb{{\mathbf b}}
\def\bh{{\mathbf h}}
\def\bg{{\mathbf g}}
\def\bla{{\mathbf \la}}
\def\bmu{\mathbf m }
\def\by{{\mathbf y}}
\def\bmu{\mbox{\boldmath $\mu$} }
\def\bsig{\mbox{\boldmath $\sigma$} }
\def\bunity{{\mathbf 1}}
\def\cA{{\cal A}}
\def\cB{{\cal B}}
\def\cC{{\cal C}}
\def\cD{{\cal D}}
\def\cF{{\cal F}}
\def\cG{{\cal G}}
\def\cH{{\cal H}}
\def\cI{{\cal I}}
\def\cL{{\cal L}}
\def\cN{{\cal N}}
\def\cM{{\cal M}}
\def\cO{{\cal O}}
\def\cR{{\cal R}}
\def\cS{{\cal S}}
\def\cT{{\cal T}}
\def\eV{{\rm eV}}

\title{Born-Infeld condensate as a possible   origin of neutrino masses and dark energy}

\author{Andrea Addazi}

\affiliation{ Dipartimento di Fisica,
 Universit\`a di L'Aquila, 67010 Coppito AQ, Italy,}
 
 \affiliation{Laboratori Nazionali del Gran Sasso (INFN), 67010 Assergi AQ, Italy,}
 
 \author {Salvatore Capozziello}

\affiliation{Dipartimento di Fisica "Ettore Pancini", Università
di Napoli {}``Federico II'', INFN Sez. di Napoli, Compl. Univ. di
Monte S. Angelo, Edificio G, Via Cinthia, I-80126, Napoli, Italy,} 

\affiliation{INFN Sez. di Napoli, Compl. Univ. di
Monte S. Angelo, Edificio G, Via Cinthia, I-80126, Napoli, Italy,}

\affiliation{Gran Sasso Science Institute 
(INFN), Viale F. Crispi 7, I-67100, L'Aquila, Italy.}

 \author {Sergei Odintsov}

\affiliation{Instituci\'o Catalana de Recerca i Estudis Avancats (ICREA), Barcelona, Spain,} 

\affiliation{Institut de Ciencies de l'Espai (IEEC-CSIC),
Campus UAB, Carrer de Can Magrans, s/n
08193 Cerdanyola del Valles, Barcelona, Spain}

\affiliation{Lab. Theor. Cosmology, Tomsk State University of Control Systems and Radioelectronics (TUSUR), 634050 Tomsk, Russia.}

\affiliation{ Tomsk State Pedagogical University, 634061 Tomsk, Russia.}

\date{\today}

\begin{abstract}

We discuss the possibility that a   Born-Infeld condensate coupled to neutrinos
can generate both neutrino masses and an effective cosmological constant. 
In particular,  an effective field theory is provided capable of dynamically realizing the neutrino superfluid phase firstly suggested by Ginzburg and Zharkov. In such a case,  neutrinos acquire a mass gap inside  the Born-Infeld ether forming  a long-range Cooper pair. Phenomenological implications of the approach are also discussed.


\end{abstract}

\maketitle

\section{Introduction}

The idea that the cosmological vacuum energy and neutrino masses have a common origin 
was suggested by several authors in the past without a final convincing implementation of it. 
It was stimulated by an intriguing coincidence of numbers: the vacuum energy is 
$\rho_{vacuum}\sim \Lambda M_{Pl}^{2}\sim (10^{-3}\, \rm eV)^{4}$
  while neutrino masses can be $m_{\nu}\sim 10^{-3}\, \rm eV$. 
Naively, the cosmological constant and neutrino masses could appear 
as disconnected phenomena: their hierarchy is  essentially given by 
\begin{equation}\frac{\Lambda}{m_{\nu}^{2}}\sim \frac{\lambda_{\nu}^{2}}{r_{H}^{2}}\sim 10^{-58}\end{equation}
where $r_{H}$ is the Hubble radius of the Universe and $\lambda_{\nu}$ the neutrino Compton wavelength. 
However, the vacuum energy is coupled with gravitational field 
by  $G_{N}\sim M_{Pl}^{-2}$, so that $\rho_{vacuum}\sim m_{\nu}^{4}$ 
and  $\Lambda=G_{N}\rho_{vacuum}\sim r_{H}^{-2}$. 
 A part these numerology, the mechanism behind the generation of  vacuum energy 
 and neutrino masses could be framed under the standard of some effective theory.
 In view of this goal,  neutrino masses could be generated
by some condensate interacting with neutrinos 
and providing a source for dark energy. This approach has been pursued in  some previous works    where the idea of neutrino mixing condensate is capable of giving rise to a dynamically evolving dark energy density \cite{Capolupo:2008rz, Capolupo:2007hy,  Capolupo:2006et,  Blasone:2004yh}. 

On the other hand, a condensate could be derived from the  Born-Infeld theory.  In particular,  the idea of an invisible Born-Infeld condensate acting as dark energy density was first suggested in Ref.\cite{Elizalde:2003ku}. It is worth  mentioning  that also the standard 
 visible electrodynamics can be extended {\it \'a l\'a} Born-Infeld so that 
it could contribute, in some sense,  to the  vacuum energy \cite{Labun:2008qq,Labun:2010xx}.

Here, we want to show how  an invisible Born-Infeld condensate can be  related to the neutrino masses and give rise to the dark energy content of the Universe. 

The old theory of a non-linear electrodynamics
was suggested as a non-linear extension of 
Maxwell electrodynamics \cite{BI}. 
The main purpose of that model was to solve the 
insidious self-energy problem of point-like charged particles. 
This model has a certain attractiveness:
it is invariant under Dirac electric-magnetic duality, 
 it predicts the existence of solitonic objects 
and it  has physical propagations
without shock waves and dichroism \cite{GZ2}. 
Even if the original attempts were not successful
for electron-self-energy, 
it turns out to be deeply connected 
to string theory.  
The Born-Infeld electrodynamics 
is the low-energy theory of open strings 
and it also re-appears as a contribution part 
of world-volume action of D3-branes as string solitons. 
So that it is well  known, in string theory,  how to  regularize 
a Born-Infeld theory in the UV limit. Otherwise problems with quantization
and unitarity would severely afflict the proposal. 
So that, in the contest of D-brane worlds or intersecting D-brane
worlds cosmology, we can envisage an invisible electrodynamics
confined on a D3-brane and gravitationally interacting in  our D-brane world. 
This picture  has important implications in 
particles physics because of non-pertubative stringy effects 
related to  exotic instantons. For instance, exotic instantons can generate
new effective operators in the low energy limit
such as an effective Majorana mass for the neutron besides neutrinos
 \cite{Addazi:2014ila,Addazi:2015ata,Addazi:2015rwa,Addazi:2015hka,Addazi:2015yna,Addazi:2015ewa}. 
 
 In this picture, the presence of open strings attached to the ordinary D-brane 
 world and the dark D3-brane can 
 generate effective interactions among 
 the Invisible Born-Infeld field and the Standard Model particles. 
 In fact, they develop a high tension, i.e. first levels of Kaluza-Klein or Regge modes
acquire a very high mass and they can be integrated out at low energy limit, 
 generating new effective operators. 

In this paper, we will argue how 
massless neutrinos can  dynamically get an effective mass term 
in the propagation inside the Born-Infeld ether. 
This inevitably leads to the formation 
neutrino-antineutrino Cooper pairs, 
leading to superfluid neutrino phase.
The superfluid neutrino phase was an old idea firstly suggested by 
Ginzburg and Zharkov \cite{GZ}. 
This idea has   phenomenological implications particularly interesting in cosmology as we are going to discuss.

The paper is organized as follows: in Sec. 2, we 
describe the formalisms of the invisible Born-Infeld
theory.  Sec. 3 is devoted to 
the neutrino condensation mechanism 
and the dynamical generation of neutrino masses.
The possibility of a neutrino superfluid phase is discussed in Sec.4
In Sec. 5,  we draw our conclusions and discuss  phenomenological implications.

\section{A Born-Infeld model for Dark energy}

Let us consider a Born-Infeld model describing an Abelian theory of electrodynamics 
with non-linear interactions 
$A_{\mu}$ formulated in 4 dimensions:
\be \label{S}
S_{BI}=-\lambda \int d^{4}x \left\{\sqrt{-{\rm det}(g_{\mu\nu}+F_{\mu\nu})}- \sqrt{-{\rm det}(g_{\mu\nu})}\right\}\,,
\ee
where 
\be\label{TDBI}
{\rm det}(g_{\mu\nu}+F_{\mu\nu})=(-g)\left[1+\frac{1}{2}F_{\mu\nu}F^{\mu\nu}+\frac{1}{16g}(F_{\mu\nu}\tilde{F}^{\mu\nu})^{2} \right]\,,
\ee
with   $g={\rm det} g_{\mu\nu}$ and $\tilde{F}_{\mu\nu}=\frac{1}{2}\epsilon_{\mu\nu\rho\sigma}F^{\rho\sigma}$.
The Lagrangian can be also rewritten as 
\be \label{L2}
\mathcal{L}=-\lambda \sqrt{1-\frac{E^{2}-B^{2}}{\lambda}-\frac{(E\cdot B)^{2}}{\lambda^{2}}}+\lambda\,,
\ee
where $E=F_{0i}$ and $B_{ij}=F_{ij}$ are the electric and magnetic fields; 
$\lambda$ is a dimensional parameter with dimensions  $[\lambda]=M^{4}$.  Let us define 
$\lambda=\mathcal{M}^{4}$ for notational  convenience for the following analysis. 
The associated energy-momentum tensor is 
\be \label{energy}
T_{\mu\nu}=\frac{1}{\sqrt{-g}}\frac{\delta S_{BI}}{\delta g_{\mu\nu}}=-\frac{\lambda}{2}\left\{\frac{g^{\mu\nu}(1+\frac{1}{2}F_{\mu\nu}F^{\mu\nu})-F_{\mu\rho}^{\rho\nu}}{\sqrt{1+\frac{1}{2}F_{\sigma\rho}F^{\sigma\rho}+\frac{1}{16g}(F_{\mu\nu}\tilde{F}^{\mu\nu})^{2} } } -g_{\mu\nu}\right\}\,,
\ee
to be compared with the standard Maxwell energy-momentum tensor
\be \label{Tmunu}
T_{\mu\nu}^{0}=\frac{1}{4}g_{\mu\nu}F_{\rho\sigma}F^{\rho\sigma}-F_{\mu\rho}F^{\rho}_{\nu}\,.
\ee
The Born-Infeld energy-momentum tensor can be rewritten 
as 
\be \label{rew2}
T_{\mu\nu}=\epsilon T_{\mu\nu}^{0}+\frac{1}{4}g_{\mu\nu}\Delta\,,
\ee
where 
\be \label{energytensor}
\epsilon=-\frac{\partial V_{eff}(\mathcal{P},\mathcal{S})}{\partial S},\,\,\,\,\,\Delta =T_{\mu}^{\mu}=-\mathcal{M}\frac{d\Phi_{eff}}{d\mathcal{M}}\,,
\ee
\be \label{Veff}
V_{eff}=-\mathcal{S}+\mathcal{M}^{4}\Phi_{eff}\left( \frac{\mathcal{S}}{\mathcal{M}^{4}},\frac{\mathcal{P}}{\mathcal{M}^{4}}\right)\,,
\ee
\be \label{Phi}
\Phi_{eff}=\frac{1}{\mathcal{M}^{4}}\mathcal{S}+(1-\sqrt{1+2\mathcal{S}/\mathcal{M}^{4}-\mathcal{P}^{2}{\mathcal{M}^{8}}})
\ee
and 
\be \label{SP}
\mathcal{S}=\frac{1}{4}F_{\mu\nu}F^{\mu\nu},\,\,\,\,\,\,\,\mathcal{P}=\frac{1}{4}\tilde{F}_{\mu\nu}F^{\mu\nu}\,.
\ee
The explicit dependence of $V_{eff}$ from $\mathcal{S},\mathcal{P}$ is 
\be \label{explicit}
V_{eff}=\mathcal{M}^{4}(1-\sqrt{1+2\mathcal{S}/\mathcal{M}^{4}-\mathcal{P}^{2}{\mathcal{M}^{8}}})\,.
\ee
In other words, the energy-momentum  tensor can assume the  form 
\be \label{Tmunu}
T_{\mu\nu}=\left(-\frac{\partial V_{eff}}{\partial \mathcal{S}} \right)(g_{\mu\nu}\mathcal{S}-F_{\mu\lambda}F^{\lambda}_{\nu})-\left( V_{eff}-\mathcal{P}\frac{\partial{V}_{eff}}{\partial \mathcal{P}}-\mathcal{S}\frac{\partial{V}_{eff}}{\partial \mathcal{S}}\right)\,,
\ee
with the identification 
\be \label{ide}
\frac{1}{4}\mathcal{M}\frac{\partial \Phi_{eff}}{\partial \mathcal{M}}=\left( V_{eff}-\mathcal{P}\frac{\partial{V}_{eff}}{\partial \mathcal{P}}-\mathcal{S}\frac{\partial{V}_{eff}}{\partial \mathcal{S}}\right)\,.
\ee
Now, we can notice that the trace-term $\Delta$ can provide a cosmological-constant-like contribution, 
while $\epsilon$ can be interpreted as a non-linear dielectric function. 
We can also rewrite  Eqs.\eqref{energytensor} as
\be \label{epsilonmeglio}
\epsilon=-\frac{\partial V_{eff}}{\partial \mathcal{S}}=\frac{1}{1+2\mathcal{S}/\mathcal{M}^{4}-\mathcal{P}^{2}/\mathcal{M}^{8}}\,,
\ee
and
\be \label{TBI}
\Delta=-4\mathcal{M}^{4}\left( 1-\sqrt{\frac{1+2\mathcal{S}/\mathcal{M}^{4}+\mathcal{S}^{2}/\mathcal{M}^{8}}{1+2\mathcal{S}/\mathcal{M}^{4}-\mathcal{P}^{2}/\mathcal{M}^{8}}}\right)\,.
\ee
Finally the two functions  $\epsilon,\Delta$ are related as
\be \label{rele}
\Delta=4\mathcal{M}^{4}\left[\epsilon(1+\mathcal{S}/\mathcal{M}^{4})-1\right]\,,
\ee
and the  energy density is 
\be \label{00}
T^{00}=\frac{\epsilon}{2}(E^{2}+B^{2})+\frac{1}{4}\Delta\,.
\ee
Let us now comment on the meaning of this result. 
First of all, the non-linear terms of Born-Infeld action
are contributing to the trace-part 
of the electromagnetic  energy-momentum tensor. 
Strictly speaking, the theory is no more conformally invariant 
at classical level. This differentiates it with respect to  the standard Maxwell  theory which is 
explicitly invariant under conformal rescaling. 
The essence of this proposal is  that the energy  density generated in such a way, which gives rise to a dynamically scaling density,   is a sort of  cosmological constant. 
However, it is worth stressing  that  the cosmological constant generated in such a model is  dynamically varying with  time. 

To be more precise,  let us assume the initial conditions
\be \label{f}
\langle F_{\mu\nu}F^{\mu\nu} \rangle_{t=t_{0}} =\langle (E^{2}-B^{2}) \rangle_{t=t_{0}}=f_{0}\neq 0\,,
\ee
\be \label{g}
\langle F_{\mu\nu}\tilde{F}^{\mu\nu} \rangle_{t=t_{0}} =\langle (E \cdot B ) \rangle_{t=t_{0}}=g_{0}\neq 0\,,
\ee
for the Born-Infeld fields in  a Fiedmann-Roberston-Walker  space-time,
 where the symbols $\langle \cdot\cdot\cdot \rangle$ indicate quantum vacuum expectation values.
Clearly, such terms can be seen as contributions for an effective cosmological constant. 
However,  we expect that the non-linear electrodynamics  leads to  quantum instabilities 
of condensates. Strictly speaking, also isotropy is expected to be dynamically violated.
As a simplifying ansatz, we can consider isotropic but 
time-varying condensates given by smoothly varying time functions
\be \label{f1}
\langle F_{\mu\nu}F^{\mu\nu} \rangle =\langle (E^{2}-B^{2}) \rangle=f(t) \,,
\ee
\be \label{g2}
\langle F_{\mu\nu}\tilde{F}^{\mu\nu} \rangle =\langle (E \cdot B ) \rangle=g(t)\sqrt{-g}\,,
\ee
where the  Born-Infeld  dynamics is expected to slowly varying with  respect to the initial conditions. 
This means that  we can assume apparently bad initial conditions [\ref{f}-\ref{g}] as a physical ansatz.

Let us now parameterize the vacuum expectation value of the operators $\langle \mathcal{S}^{n}\mathcal{P}^{m} \rangle$, $\langle E_{i}^{2}\rangle$ and  $\langle B_{i}^{2}\rangle$ as
\be \label{effes}
\langle \mathcal{S}^{n}\mathcal{P}^{m} \rangle=\alpha^{n}\beta^{m}\,,
\ee
\be \label{EB}
\langle E_{i}^{2}\rangle=\langle B_{i}^{2}\rangle=\epsilon(t)\,,
\ee
following  the equipartition principle. The condition (\ref{effes}) parameterizes the quantum condensate contribution, while (\ref{EB}) are the  energy contributions of the classical fields. 
This implies the following density and pressure relations  
\be \label{rhoBI}
\rho_{BI}=\frac{1}{2}\mathcal{M}^{4}\left(\frac{1+\frac{\alpha}{2}-\frac{\alpha_{t}}{4}}{\sqrt{1+\frac{\alpha}{2}-\frac{\beta^{2}}{2}}}-1 \right)\,,
\ee
\be \label{pBI}
p_{BI}=-\frac{1}{2}\mathcal{M}^{4}\left(\frac{1+\frac{\alpha}{2}-\frac{\alpha_{s}}{4}}{\sqrt{1+\frac{\alpha}{2}-\frac{\beta^{2}}{2}}}-1 \right)\,,
\ee
where 
\be \label{alphats}
\alpha_{t}=\alpha-4\epsilon, \,\,\,\,\,\,\alpha_{s}=\alpha+\frac{4}{3}\epsilon\,.
\ee
 As remarked in Ref.\cite{Elizalde:2003ku},  the parameter $\alpha$ has a quantum origin.  It gives the  parameterization of the quantum vacuum expectation value of the correlator (\ref{effes}). 
In fact, since the electric field is perpendicular to the magnetic field, 
$\langle F \tilde{F}\rangle=\langle E\cdot B \rangle=0$ is expected 
at purely classical level, i.e. $\alpha=0$ in the expression (\ref{effes}).

Now, the energy-momentum tensor of the Born-Infeld  field has to be conserved according to the Bianchi identities, that is:
\be \label{cons}
\nabla_{\mu}T^{\mu\nu}_{BI}=0\,\,\,\,\rightarrow \nabla_{\mu}\left[-\frac{\lambda}{2}\left\{\frac{g^{\mu\nu}(1+\frac{1}{2}F_{\mu\nu}F^{\mu\nu})-F_{\mu\rho}^{\rho\nu}}{\sqrt{1+\frac{1}{2}F_{\sigma\rho}F^{\sigma\rho}+\frac{1}{16g}(F_{\mu\nu}\tilde{F}^{\mu\nu})^{2} } } -g_{\mu\nu}\right\}\right]=0\,.
\ee
Under the  assumption of the Friedmann-Robertson-Walker background, Eq.\eqref{cons} reduces to  the simple form 
\be \label{simplef}
\dot{\rho}_{BI}+3H(\rho_{BI}+p_{BI})=0\,,
\ee
where $H=\dot{a}/a$. 
From Eqs.\eqref{rhoBI}, \eqref{pBI}, 
and expanding up to the first order the  $\epsilon$ parameter, we obtain
\be \label{w}
w=-1+\frac{128\epsilon}{96\epsilon+3\beta^{3}}\,.
\ee
This assumption has  physical meaning  in the limit of $\alpha>>\epsilon$,
 that means that  the quantum condensate dominates on the classical field contribution. 
In other words, the equation of state  $w(t)$ will evolve in time in the limit of quantum condensate dominance.

\section{Neutrino condensation in  Born-Infeld electrodynamics}

The above considerations can be straightforwardly related to the phenomenon of neutrino condensation.
Let us suppose that neutrinos interact with the invisible electrodynamics according to the following Lagrangian
\be \label{Lpsi}
\mathcal{L}_{\nu}=i\bar{\nu}\gamma_{\mu}\partial^{\mu}\nu+g_{\nu}A_{\mu}\bar{\nu}\gamma^{\mu}\nu\,,
\ee
or, alternatively,  
\be \label{Lpsi}
\mathcal{L}_{\nu}=i\bar{\nu}_{L}\gamma_{\mu}\partial^{\mu}\nu_{L}+g_{\nu}A_{\mu}\bar{\nu}_{L}\gamma^{\mu}\nu_{L}\,.
\ee

Two neutrino particles with opposite chirality with an attractive interaction between them 
form a Cooper pair. At tree level, neutrinos and antineutrinos exchange a Born-Infeld 
dark photon with an effective attractive potential, which induces neutrino condensation.


\begin{figure}[htb] \label{FIGURA1}
\begin{center}
\caption{ a) Photon interaction with the background Born-Infeld condensate through the Euler-Heisenberg interaction term.  b)
Neutrino interaction with the superfluid condensate through an effective four-fermion interaction induced by the Born-Infeld interaction.} 
\includegraphics[scale=0.1]{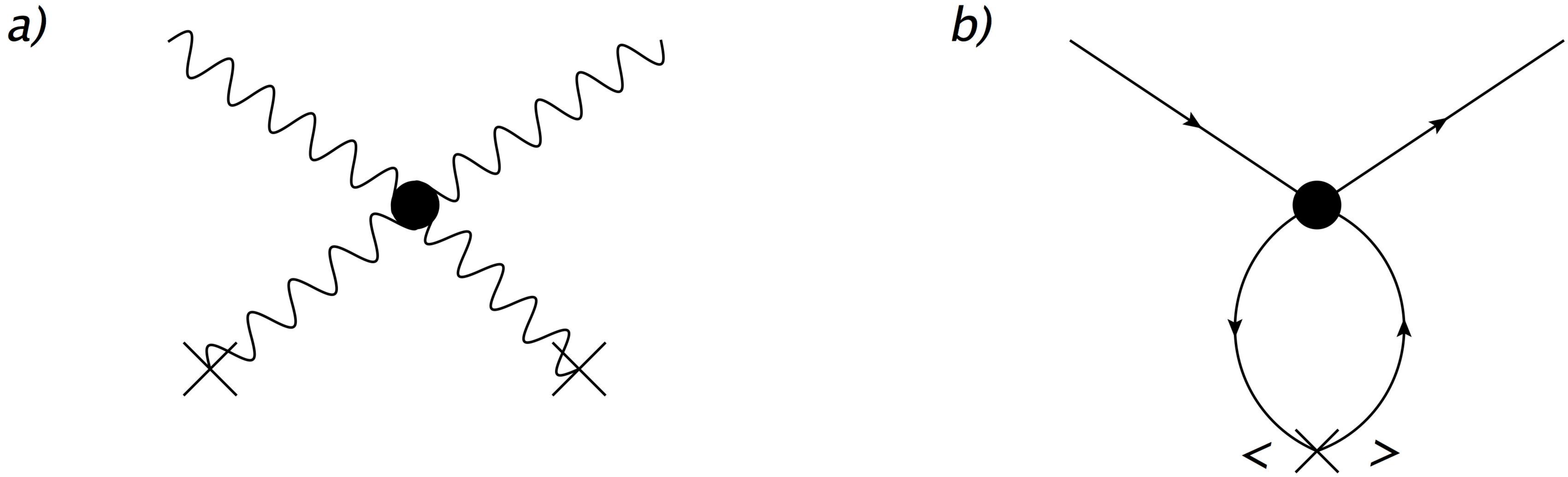}  
\end{center}
\vspace{-3mm}
\end{figure}

The condensation process can only happen in non-relativistic regime, 
with the  four-neutrinos effective interaction given by the Hamiltonian
\be \label{Hqq}
H_{int}=-\sum_{k,\sigma} V_{kk}c_{k',\sigma}^{\dagger}c_{k,\sigma}^{\dagger}c_{k,\sigma}c_{k',\sigma}\,,
\ee
$$V_{kk}=\frac{4\pi^{2}g_{\nu}^{2}}{q^{2}+V_{eff}''}\,,$$
On the other hand, the propagating dark Born-Infeld photon 
in the constant background gets an effective mass
which, for small momentum transfers $q<<\mathcal{M}$,
is $m^{\gamma}_{eff}\simeq \mathcal{M}+O(q^{2})$.
This because of the higher-derivative Euler-Heisenberg interactions 
among propagating photons and background field photons. 
So that, 
\be \label{Veff}
V_{eff}''=\mathcal{M}^{2}+O(q^{4}),\,\,\,\,V_{kk'}\simeq \frac{2\pi^{2}g_{\nu}^{2}}{2\mathcal{M}^{2}}\,,
\ee
 As a consequence, an effective four fermion effective interaction is generated with
a effective Fermi-like coupling $\bar{G}_{F}\simeq -2\pi^{2}g_{\nu}^{2}/\mathcal{M}^{2}$.
In the quantum field theory formalism, the corresponding interaction term 
has the form $\bar{G}_{F}\bar{\nu}\gamma_{\mu}\nu\gamma^{\mu}\bar{\nu}\nu$.
On the other hand, a neutrino propagating in the Born-Infeld background gets the
effective mass term $m_{eff}^{\nu}\simeq g_{\nu}\mathcal{M}$.
However, there is also an extra effective mass contribution for neutrino 
coming from the effective occupation number of cold neutrino,
which is parametrized by a chemical potential $\rho$ and which we will discuss later. 

As a consequence, neutrinos which are cooled enough by the background
($E<\mathcal{M}$) are described by  the  non-relativistic Lagrangian 
\be \label{Form}
\mathcal{L}=\sum_{\alpha=\uparrow,\downarrow}\nu^{\dagger}_{\alpha}\left(i\frac{\partial}{\partial t}-\frac{\nabla^{2}}{2m^{\nu}_{eff}}+\mu \right)\nu_{\alpha}-\bar{G}_{F}\nu^{\dagger}_{\uparrow}\nu_{\downarrow}^{\dagger}\nu_{\downarrow} \nu_{\uparrow}\,,
\ee
 which leads to the expectation value (in the Born-Infeld medium) 
\be \label{nunubar}
\langle \bar{\nu} \nu \rangle=\mu/\bar{G}_{F}=\mu \mathcal{M}^{2}/(2\pi^{2}g_{\nu}^{2})\,,
\ee

 It is convenient to work in the matrix notation $\Psi=(\nu_{\uparrow},\nu^{\dagger}_{\downarrow})^{T}$, $\Psi^{\dagger}=(\nu_{\uparrow}^{\dagger},\nu_{\downarrow})$,
with a partition function $Z=\int \mathcal{D}\Psi^{\dagger}\mathcal{D}\Psi e^{i\int d^{4}x\mathcal{L}}$.
Considering a scalar field  $\Phi$,  we can rewrite the four-fermion interaction term as 
\be \label{gpsipsi}
\bar{G}_{F}|\nu_{\downarrow}\nu_{\uparrow}|^{2}=\frac{1}{\bar{G}_{F}}|\Phi-\bar{G}_{F}\nu_{\downarrow}\nu_{\uparrow}|^{2}-\frac{1}{\bar{G}_{F}}|\Phi|^{2}+\Phi^{\dagger}\nu_{\downarrow}\nu_{\uparrow}+h.c.\,,
\ee
However, we are interested to the combination $\Phi=\bar{G}_{F}\nu_{\uparrow}\nu_{\downarrow}$,
which renders null the first term. 
The partition function can be rewritten in $\Phi,\Phi^{\dagger}$ variables as 
\be \label{Zpdi}
Z=\int \mathcal{D}\Psi^{\dagger}\mathcal{D}\Psi \int \mathcal{D}\Phi^{\dagger} \mathcal{D}\Phi e^{iS[\Psi,\Psi^{\dagger},\Phi,\Phi^{\dagger}]}\,,
\ee
\be \label{Spspsp}
S[\Psi,\bar{\Psi}^{\dagger},\Phi,\Phi^{\dagger}]=\int d^{4}x\left[ \Psi^{\dagger} \left( \begin{array}{cc} i\partial_{t}-\frac{\nabla^{2}}{2m_{eff}^{\nu}}+\mu & -\Phi
\ \\  -\Phi^{\dagger} & i\partial_{t}-\frac{\nabla^{2}}{2m_{eff}^{\nu}}-\mu    \ \\
\end{array} 
  \right)\Psi+\frac{1}{\bar{G}_{F}}|\Phi|^{2}\right]\,.
\ee
 Now, using the Grassmanian integral 
\be \label{Grass}
\int \mathcal{D}\Psi \mathcal{D}\Psi^{\dagger}e^{-\Psi^{\dagger}M\Psi}={\rm det}\, M\,,
\ee
 the partition function can be rewritten as 
\be \label{rewritten}
Z=\int \mathcal{D}\Phi^{\dagger}\mathcal{D}\Phi \,\mathcal{O} \, e^{\frac{i}{\bar{G}_{F}}\int d^{4}x|\Phi|^{2}}\,,
\ee
and 
$$\mathcal{O}=\left( \begin{array}{cc} i\partial_{t}+\frac{\nabla^{2}}{2m_{eff}^{\nu}}+\mu & -\Phi
\ \\  -\Phi^{\dagger} & i\partial_{t}-\frac{\nabla^{2}}{2m_{eff}^{\nu}}-\mu    \ \\
\end{array} 
  \right)\,,$$
 which is equivalent to 
\be \label{equiv}
Z=\int \mathcal{D}\Phi^{\dagger}\mathcal{D}\Phi e^{iS[\Phi,\Phi^{\dagger}]}\, ,
\ee
$$S[\Phi,\Phi^{\dagger}]=\int d^{4}x \left[\frac{i}{\bar{G}_{F}}|\Phi|^{2}+\int \frac{d^{4}k}{(2\pi)^{4}}{\rm tr}
 \left( \begin{array}{cc} k^{0}-\frac{|{\bf k}|^{2}}{2m_{eff}^{\nu}}+\mu & -\Phi
\ \\  -\Phi^{\dagger} & k^{0}+\frac{|{\bf k}|^{2}}{2m_{eff}^{\nu}+\mu}-\mu    \ \\
\end{array} \right)   \right]\, .$$
 At this point, we can try the extremal solution of the action 
in the ground state of $\Phi$, 
i.e. $\Phi\rightarrow \langle \Phi \rangle$
and $\delta S[\langle \Phi \rangle, \langle \Phi^{\dagger} \rangle]=0$. 
One can show that the action is minimized 
by the condition
\be \label{Phi}
0=\frac{i}{\bar{G}_{F}}\langle \Phi \rangle +\int \frac{d^{4}k}{(2\pi)^{4}}\frac{1}{(k^{0})^{2}-(|{\bf k}/2m_{eff}^{\nu}-\mu)^{2}-|\langle \Phi \rangle|^{2}+i\epsilon}{\rm tr} \,\mathcal{O}_{2}\, ,
\ee
where 
\be \label{O2}
\mathcal{O}_{2}=\left(\begin{array}{cc} -\langle \Phi \rangle & 0
\ \\  k^{0}-|{\bf k}|^{2}/2m_{eff}^{\nu}+\mu & 0    \ \\
\end{array} \right)\, .
\ee
 Eq.(\ref{O2}) explicitly provides the fermion propagator 
 with poles
\be \label{fermion}
E_{\pm}=\pm \sqrt{(|{\bf k}|^{2}/2m_{eff}^{\nu}-\mu)^{2}+|\langle \Phi \rangle|^{2}}\, ,
\ee

In the Hamiltonian approach, the Hamiltonian of Cooper pairs reads as
\be \label{Hqq}
H=2\sum_{k}\epsilon_{k}b_{k}^{\dagger}b_{k}-\sum_{k\neq k'}V_{kk'}b^{\dagger}_{k'}b_{k}\,,
\ee
where $\epsilon_{k}=k^{2}/2m_{eff}^{\nu}-\mu+V_{kk'}/2$ 
and
 pairing operators $b,b^{\dagger}$ create and destroy Cooper pairs of neutrinos
on the ground state $|\Psi_{0}\rangle$:
$$b_{k}^{\dagger}=c^{\dagger}_{k}c_{-k}^{\dagger},\,\,\,\,b_{k}=c_{-k}c_{k}\,.$$

The formation of a Cooper pair condensate 
$\langle \bar{\nu}_{R} \nu_{L}\rangle$
breaks the chiral symmetry as 
$$SU(2)_{L}\times SU(2)_{R}\rightarrow SU(2)_{V}$$
generating the scalar Nambu-Goldstone bosons. 
While neutrinos with energies $E<\mathcal{M}$ will undergo to condensation, 
neutrinos with $E>>\mathcal{M}$ are practically unbounded,
but they get a mass term as 
\be \label{nunuLR}
-\bar{G}_{N}\langle \bar{\nu}_{L} \nu_{R} \rangle \bar{\nu}_{L} \nu_{R}\rightarrow    \mu\, \bar{\nu}_{L} \nu_{R}\, .
\ee
The $\mu$ parameter can be  interpreted as the finite density parameter of the condensate or  the Fermi energy level,
which is $\mu \sim \mathcal{M}$. 

The propagator of neutrinos in the Born-Infeld medium is defined 
as 
\be \label{SF}
S_{F}(x,y)=\langle T\left\{  \bar{\nu}\nu\right\} \rangle,\,
\ee
where $\langle |...|\rangle$ denotes the correlator in the 
non-trivial Born-Infeld vacuum. 
 As argumented in Ref.\cite{Labun:2008qq,Labun:2010xx}, 
the  effective formula for fermion propagator inside the invisible Born-Infeld ether is
\be \label{free}
g_{\nu}\mathcal{M}\frac{\partial V_{eff}(x,\mathcal{M})}{\partial \mathcal{M}}=ig_{\nu}\mathcal{M}\,  {\rm lim}_{\epsilon\rightarrow 0}{\rm tr}[\Delta S(\epsilon,x)]\,,
\ee
where 
\be \label{Delta}
\Delta S(\epsilon,x)=S_{F}(x+\epsilon,x-\epsilon,\mathcal{M})-S_{F}^{0}\,,\ee
and  $S_{F}^{0}$ is the free-field propagator. 
In Eq.\ref{free}, the non-linear correction to the free-propagator are 
contained in the term $\mathcal{M}\partial V_{eff}(x,\mathcal{M})/\partial \mathcal{M}$.
Eq.\ref{free} implies 
\be \label{mass}
g_{\nu}\mathcal{M}(iS_{F}(x,x)-i S^{0}_{F}(x,x))=g_{\nu}\mathcal{M} \frac{\partial V_{eff}(x,\mathcal{M})}{\partial \mathcal{M}}
 \rightarrow -g_{\nu}\mathcal{M}  \bar{\nu}\nu  \,.\ee
where we used the Wick theorem $\langle :\nu \bar{\nu}:\rangle=S_{F}(x,x)-S^{0}_{F}(x,x)$, Eq.(\ref{Delta}) and $:\nu \bar{\nu}: =-\bar{\nu}\nu$.
So that a tachyonic mass term for condensing neutrinos is generated, in agreement with arguments given above. 

Finally, let us comment on the possible Pontecorvo mass matrix 
necessary for neutrino oscillations. 
We can easily generalize the one-neutrino structure
in Lagrangian (\ref{Lpsi}) with 
\be \label{Lpsi2}
\mathcal{L}_{\nu}=i\bar{\nu}_{f}\gamma_{\mu}\partial^{\mu}\nu_{f}+G_{ff'\nu}A_{\mu}\bar{\nu}_{f}\gamma^{\mu}\nu_{f'}\,,
\ee
where $f=1,2,3$ is the flavor index of neutrinos and $G_{ff'\nu}$ is a flavor mixing matrix. 
From this Lagrangian we can generate a Pontecorvo matrix $\mathcal{M}_{ff'}=G_{ff'\nu}\langle \bar{\nu}\nu \rangle$.

\section{Neutrino superfluid phase}
 We will  now discuss the  intriguing possibility of a  superfluid phase
induced by Cooper pairs of neutrinos. This fact could have interesting cosmological consequences due to the cosmic neutrino background.
 As in the standard  superfluidity theory, the problem can be treated from two prospectives:  considering the Ginzburg-Landau model as an effective theory 
describing the Cooper pair as a scalar field \cite{Landau};  considering the Bogolubov approach as an effective theory of interacting fermions (neutrinos)
\cite{Bogoliubov}. 

In the Ginzburg-Landau approach, 
one consider a non-relativistic effective field theory of Cooper scalar fields in a finite density $\rho$.
The effective Lagrangian is 
\be \label{effL}
\mathcal{L}=i\Phi^{\dagger}\partial_{0}\Phi-\frac{1}{2m_{eff}^{\nu}}\partial_{i} \Phi^{\dagger}\Phi-g^{2}(\Phi^{\dagger}\Phi-\rho)^{2}\,,
\ee
where $m_{eff}=g_{\nu}\mathcal{M}$ and  $g$ is the Ginzburg-Landau adimensional effective coupling. 
Confronting Eq.\ref{effL} with Eq.\ref{nunubar}, the $\mu$ parameter is equal to the finite density parameter $\rho$. 
In polar variables, 
$\Phi=\sqrt{\tilde{\rho}}e^{i\theta}$, 
where $\sqrt{\tilde{\rho}}=\sqrt{\rho}+a$.
The Lagrangian \eqref{effL}
becomes 
\be \label{becL}
\mathcal{L}=-\tilde{\rho}\, \partial_{0}\theta-\frac{1}{m_{eff}^{\nu}}\left[\frac{1}{\tilde{\rho}}(\partial_{i}\tilde{\rho})^{2}+\tilde{\rho} (\partial_{i}\theta )^{2}\right]-g^{2}(\tilde{\rho}-\rho)^{2}\,,
\ee
that is, by developing and expanding, 
\be
\mathcal{L}=-2\sqrt{\rho}\,a\dot{\theta}-\frac{\rho}{2m_{eff}^{\nu}}(\partial_{i}\theta)^{2}-\frac{1}{2m_{eff}^{\nu}}(\partial_{i}a)^{2}-4g^{2}\rho a^{2}+...
\ee
Integrating out $a$, we obtain 
\be \label{LLL}
\mathcal{L}=\rho \dot{\theta}\frac{1}{4g^{2}\rho-(1/2m_{eff}^{\nu})\partial_{i}^{2}}\dot{\theta}-\frac{\rho}{2m_{eff}^{\nu}}(\partial_{i}\theta)^{2}\,.
\ee
This implies the presence of gapless Bogolubov modes 
propagating with a dispersion relation of the form 
\be \label{omega}
\omega^{2}=\left(\frac{2g^{2}\rho}{m_{eff}^{\nu}}\right)k_{i}^{2}\,.
\ee
Of course the low energy limit of this theory is simply reduced to the  Lagrangian 
of a Nambu-Goldstone boson of the form
\be \label{LLLL}
\mathcal{L}=\frac{1}{4g^{2}}\partial_{\mu}\theta \partial^{\mu}\theta\,.
\ee

 Now let us consider the  Bogolubov approach.
The effective Hamiltonian is 
\be \label{effective}
\mathcal{H}=\int d^{3}x\left[\nu_{\alpha}^{\dagger}(r,t)\left(-\frac{1}{m_{eff}^{\nu}}\nabla^{2}-E_{F}\right)\nu_{\alpha}(r,t)
+O(r,t)\nu_{\uparrow}(r,t)\nu_{\downarrow}(r,t)^{\dagger}+h.c.\right]\,,
\ee
where the Fermi energy $E_{F}$ is the chemical potential of the system. 
 $O(r,t)$ is the order field, which reads
\be \label{O}
O_{\alpha\dot\beta}(r,t)=\bar{G}_{N}\langle\nu_{\alpha}\bar{\nu}_{\dot\beta} \rangle\,.
\ee
Comparing Eq.\ref{effective} and Eq.\ref{equiv}, the order field is a mean field of the quantum field operator $\Phi$,
while the Fermi energy scale $E_{F}$ coincide with $\mu$. 
$\nu(r,t)$ is the non-relativistic neutrino wave function. 
The equations of motion  are given by  the standard Bogolubov-de Gennes matrix equation:
\be \label{EoM}
i\partial_{t}\left( \begin{array}{c} \nu_{\uparrow} 
\ \\ \nu_{\downarrow}^{\dagger}  \ \\
\end{array} \right)=     \left( \begin{array}{cc} -\frac{1}{2m_{eff}^{\nu}}\nabla^{2}-E_{F} & O(r,t)
\ \\  O^{\dagger}(r,t) & \frac{1}{2m_{eff}^{\nu}}\nabla^{2}+E_{F}    \ \\
\end{array} \right)
\left( \begin{array}{c} \nu_{\uparrow} 
\ \\ \nu_{\downarrow}^{\dagger}  \ \\
\end{array} \right)\,.
\ee
Clearly if $O=0$, the system of equations of motion decouples 
and the dispersion relations are reduced to 
particles-antiparticles being  $E_{\pm}=\pm (p^{2}/2m_{eff}^{\nu}-E_{F})$.
If  $O\neq 0$, 
quasiparticles get gap
as $E^{2}=E_{\pm}^{2}+|O|^{2}$.
and  quasiparticles  wave functions are
\be \label{quasi}
\left( \begin{array}{c} \phi_{p}
\ \\ \phi_{h}  \ \\
\end{array} \right)=C_{1}e^{-iE_{+}t+ip_{+}\cdot r}+C_{2}e^{-iE_{-}t+ip_{-}\cdot r}	\,,
\ee
where 
$$C_{1}=c_{1}\left( \begin{array}{c} a_{+}e^{i\delta}
\ \\ a_{-}e^{-i\delta}  \ \\
\end{array} \right)\,,$$
$$C_{2}=c_{2}\left( \begin{array}{c} a_{-}e^{i\delta}
\ \\ a_{+}e^{-i\delta}   \ \\
\end{array} \right)\,,$$
$$a_{\pm}=\sqrt{\frac{1}{2}(1\pm \frac{1}{|E_{\pm}|}\sqrt{E_{\pm}^{2}-|O|^{2}})}\,,$$
$$p_{\pm}^{2}/2m_{eff}^{\nu}=E_{F}\pm \sqrt{E_{\pm}^{2}-|O|^{2}}\,,$$
where $c_{1,2}$ are constants and  $\delta$ is the phase of 
the complex order operator; 
$\phi_{p,h}$ can be seen as particle and hole excitations.

\section{Conclusions and outlooks}

In this paper, we explored a mechanism 
generating neutrino masses and 
dark energy in the  unified framework given by the Born-Infeld theory. 
In particular, we showed how an invisible non-linear 
electrodynamics coupled to neutrinos can 
provide both a Dirac neutrino mass term
and a dark energy contribution. 
This is also related to an exotic neutrino superfluid state 
firstly conjectured in \cite{GZ}.

We find similar phenomenological implications of the 
model recently suggested by Dvali and Fucke \cite{Dvali:2016uhn}, 
even if our approach is  different in origin and formulation.
They suggested that the neutrino mass is somehow 
generated by a gravitational $\theta$-term condensate,
by a  gravitational condensation mechanism. 
Inspired by their results, we can resume the following
 implications for  our model:

- In the minimal version of this model,
the neutrino is a Dirac not a Majorana particle. 
Of course the model can be extended with Majorana-mass generating 
mechanisms \cite{ss1,ss2,ss3,ss4}, having in mind see-saw mechanisms and leptogenesis \cite{ss5}. 

- The cosmological neutrino mass bound vanishes: neutrinos
are massless until the phase transition at redshift $z\sim 1$
corresponds to $T\sim meV$.

- The flavor ratio of neutrinos could be constrained 
by IceCube data \cite{ice}; KATRIN experiment \cite{katrin}
could detect an overdense cosmic neutrino fluid in 
our Galaxy. 

- In conclusion, neutrinos and antineutrinos annihilates into Nambu-Goldstone bosons 
leaving part of relic cosmological neutrinos
while contributing to at least a part of cold-dark matter.
As argued, they are converted from hot dark matter to cold dark matter 
during the phase  transition. 

We also would like to point out  another possible connection with a model
proposed by Berezhiani and Khoury where  a superfluid dark matter 
can have Bogolubov modes \cite{Berezhiani:2015pia,Berezhiani:2015bqa}
interacting with the gravitational field and modifying 
the Newtonian potential in our galaxy. 
This can lead to a hybrid MOND/CDM model, 
with interesting  consequences for  phenomenology in astrophysics 
and cosmology to be explore in future. 

In our model, we also derived that the neutrino masses are slowly 
running with the Invisible Born-Infeld field. 
We think that phenomenological implications of this prediction 
will deserve future  investigations which go beyond the purposes of this paper.

A  further open issue  remains the 
stabilization of the  cosmological term with respect to the  
radiative corrections.  This problem could be 
related to  quantum gravity and then addressed in some future unified theory.

An alternative model to our approach has been  recently suggested:
the dark energy could be generated by an invisible Yang-Mills
or Invisible QCD condensates
  \cite{Dona:2015xia,Addazi:2016sot,Alexander:2016xbm}
In this case, the generation 
of a neutrino mass gap is similarly possible 
but it seems ruled out by observations of unconfined neutrinos. 
As final remark, it is worth saying that nonlinear electrodynamics could contain the whole budget of non-gravitational background of the Universe comprising neutrinos, radiation and dark energy in the form of a condensate. Further theoretical investigations and experimental evidences are necessary in this perspective.

\begin{acknowledgments} 

AA would like to thank Gia Dvali for valuable comments and discussions and 
Ludwig Maximilan University in Munich  for hospitality during the preparation of this paper.
The work of AA  is supported in part by the MIUR research
grant "Theoretical Astroparticle Physics" PRIN 2012CPPYP7
and by SdC Progetto speciale
Multiasse {\it La Societ\'a della Conoscenza in Abruzzo} PO FSE Abruzzo 2007-2013.
SC is supported in part by INFN {\it (iniziative specifiche} QGSKY and TEONGRAV).
SDO is supported in part by MINECO (Spain), project FIS2013-44881.
This article is partly based upon work from COST Action CA15117 (Cosmology and Astrophysics Network for
 Theoretical Advances and Training Actions), supported by COST (European Cooperation in Science and Technology).
\end{acknowledgments}

\end{document}